\documentclass{ceurart}

\usepackage[inline]{enumitem}
\usepackage{physics}
\usepackage{graphicx}
\usepackage{caption}
\usepackage{subcaption}
\usepackage{booktabs}

\begin{document}

\copyrightyear{2025}
\copyrightclause{Copyright for this paper by its authors. Use permitted under Creative Commons License Attribution 4.0 International (CC BY 4.0).}
\conference{BigHPC2025: Special Track on Big Data and High-Performance Computing, co-located with the 4\textsuperscript{th} Italian Conference on Big Data and Data Science, ITADATA2025, September 9 -- 11, 2025, Torino, Italy.}
\title{Exploring an implementation of quantum learning pipeline for support vector machines}

\author[1]{Mario Bifulco}[email=mario.bifulco@unito.it,url=https://github.com/TheFlonet/End2EndQuantumSVM]
\cormark[1]
\author[1]{Luca Roversi}[orcid=0000-0002-1871-6109,email=luca.roversi@unito.it,url=https://www.di.unito.it/~rover/]
\address[1]{Università degli studi di Torino, Dipartimento di Informatica, Corso Svizzera 185 - 10149 Torino}
\cortext[1]{Corresponding author.}

\begin{abstract}
  This work presents a fully quantum approach to support vector machine (SVM) learning by integrating gate-based quantum kernel methods with quantum annealing-based optimization.
  We explore the construction of quantum kernels using various feature maps and qubit configurations, evaluating their suitability through Kernel-Target Alignment (KTA).
  The SVM dual problem is reformulated as a Quadratic Unconstrained Binary Optimization (QUBO) problem, enabling its solution via quantum annealers.
  Our experiments demonstrate that a high degree of alignment in the kernel and an appropriate regularization parameter lead to competitive performance, with the best model achieving an F1-score of 90\%.
  These results highlight the feasibility of an end-to-end quantum learning pipeline and the potential of hybrid quantum architectures in quantum high-performance computing (QHPC) contexts.
\end{abstract}


\maketitle

%---------------------
\section{Introduction}

Quantum computing ultimately aims to expand the boundaries of what is currently considered efficiently computable.
Among the various fields where such advancements appear promising is machine learning.

When discussing quantum computing, it is essential to specify the computational paradigm being considered.
\emph{Gate-based quantum computing} refers to a universal quantum computing model, typically implemented using superconducting circuits, where quantum circuits are programmed analogously to how logic gates define classical circuits.
\emph{Annealing-based quantum computing}, on the other hand, refers to a non-universal subset of adiabatic quantum computing, primarily designed for solving optimization problems.

In the context of quantum machine learning, support vector machines (SVMs) have been studied under both computational paradigms\cite{qsvm_qa, qsvm_gqc}.
Within SVMs, we can distinguish two principal components that contribute to model training:
\begin{enumerate*}
  \item The \emph{kernel method}, which enables the mapping of examples into a higher-dimensional feature space, thereby increasing the likelihood of linear separability, as per Cover's theorem\cite{th_cover};
  \item The \emph{optimization problem}, which identifies the model parameters used for inference on new data instances.
\end{enumerate*}

In the literature, the term \emph{Quantum Support Vector Machine (QSVM)} is used with two distinct meanings.
In the quantum annealing framework, QSVM refers to the use of classical kernels combined with a reformulation of the optimization problem to be solved via quantum annealing.
In the gate-based approach, by contrast, quantum computing is employed to compute the kernels, while the optimization is handled by classical processors.

This work aims to unify the aforementioned approaches to construct a fully quantum learning pipeline for support vector machines using a kernel based on quantum gates and a model optimization process based on quantum annealing.
Moreover, the use of different types of QPUs enables experimentation within the domain of Quantum High Performance Computing (QHPC).
In this domain, traditional CPU and GPU systems can collaborate with various QPUs—regardless of their underlying architecture—with the goal of addressing computationally intensive problems.

To evaluate the proposed pipeline, we employed a subset of the Breast Cancer dataset~\cite{dataset}.
The samples were randomly selected by iterating over the first 10{,}000 prime numbers as seeds, in order to identify the one that best preserved the original statistical distribution.
Breast Cancer dataset was chosen due to its well-established use in benchmarking classification algorithms and its balanced representation of clinically relevant features, which make it particularly suitable for evaluating the robustness and generalizability of machine learning pipelines.

%---------------------
\section{A short recap about gate-based generated Quantum Kernels}

This section briefly recalls the main steps to construct a gate-based quantum circuit that implements a quantum kernel.

Let $V$ be a set of input vectors.
\begin{enumerate}
  \item Each $\vec{x}\in V$ is interpreted as a quantum state $\ket{\phi(\vec{x})}$ by means of a quantum circuit $\phi(\vec{x})$;
  \item For each pair $(\vec{x}_i, \vec{x}_j)$,  the quantum circuit $\phi(\vec{x}_i);(\phi(\vec{x}_j))^{-1}$ is constructed. It pipelines $\phi(\vec{x}_i)$, building the state $\ket{\phi(\vec{x}_i)}$, with the inverse $(\phi(\vec{x}_j))^{-1}$ of $\phi(\vec{x}_j)$, building the state $\ket{\phi(\vec{x}_j)}$;
  \item The overlap between the states $\ket{\phi(\vec{x}_i)}$ and $\ket{\phi(\vec{x}_j)}$ is estimated by measuring the probability of the resulting state collapsing to the all-zero state, i.e., computing $|\langle\phi(\vec{x}_i)|\phi(\vec{x}_j)\rangle|^2$.
\end{enumerate}
This procedure leads to define a kernel matrix $K$ such that each entry $K_{ij}$ corresponds to the similarity between  $\vec{x}_i$ and $\vec{x}_j$ as inferred from their representation as quantum states $\ket{\phi(\vec{x}_i)}$ and $\ket{\phi(\vec{x}_j)}$.
Intuitively, the advantage of constructing $K$ as above lies in the ability of quantum kernels to exploit qubit entanglement and state superposition to compress the number of qubits required. Adding a single qubit to the feature map $\phi$ doubles the dimensionality of the space to which the input $x_i$ is mapped.

%---------------------
\section{Quantum Annealing Support Vector Machine}

Annealing-based quantum computing naturally addresses quadratic optimization problems, particularly those formulated as QUBO (Quadratic Unconstrained Binary Optimization) instances.
The dual formulation of the support vector machine (SVM) problem \cite{svm} leads to a multivariate quadratic polynomial, which serves as the objective function to be optimized.
By incorporating constraints into the objective via penalty terms and converting continuous variables into their binary expansion, the problem can be reformulated as a QUBO model.

Constraint incorporation is typically achieved through Lagrangian relaxation \cite{lagrangian_relaxation}.
However, the binarization process is less straightforward due to the continuous nature of the optimization variables, which lie in $\mathbb{R}$.
On one hand, integer variables can be automatically encoded in binary format, provided that an upper bound is defined for each variable.
On the other hand, the binarization of real-valued variables incurs significantly more overhead.

Empirically, we found that treating all optimization variables as integers yields a degradation in model performance of less than one percentage point.
This trade-off led us to favor an approximated yet more hardware-efficient formulation during model generation.

An additional hyperparameter of practical importance is a \emph{regularization parameter} $C$. It controls the trade-off between maximizing the classification margin and minimizing classification errors.
In this context, the margin refers to the distance between the separating hyperplane and the closest data points from each class.

Higher values of $C$ emphasize correct classification of training examples, whereas lower values allow for a softer margin with more tolerance to misclassified points.

%---------------------
\section{Results}

Both the design of a kernel and the optimization of the SVM problem require the selection of hyperparameters prior to training.
Section \ref{sec:qkernel} presents our investigation into the parametrization of the construction of a quantum kernel in the context of gate-based architectures, with respect to the chosen dataset.
Section \ref{sec:qasvm} reports on a set of experiments aimed at selecting the parameter $C$ in the context of quantum-anneal based architectures, assuming no kernel is employed during training.

Once the optimal hyperparameters for the individual components of the problem have been independently identified, Section \ref{sec:qkernel_and_qasvm} presents the results of the end-to-end learning pipeline for fully quantum SVMs—namely, those employing a kernel generated via quantum gates and optimization performed through quantum annealing.

\subsection{Finding a good Quantum Kernel}\label{sec:qkernel}

\begin{figure}
  \centering
  \begin{minipage}[b]{0.49\textwidth}
    \centering
    \includegraphics[width=\linewidth]{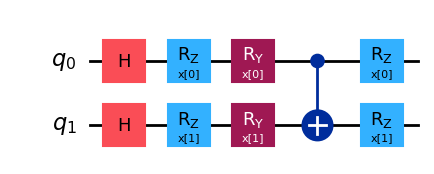}
    \subcaption{\texttt{SU2HR} feature map\label{fig:map_a}}
  \end{minipage}
  \hfill
  \begin{minipage}[b]{0.49\textwidth}
    \centering
    \includegraphics[width=\linewidth]{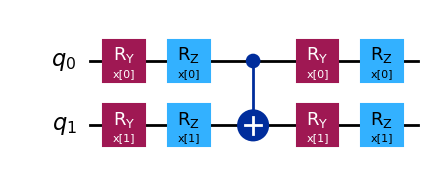}
    \subcaption{\texttt{SU2RR} feature map\label{fig:map_b}}
  \end{minipage}

  \begin{minipage}[b]{\textwidth}
    \centering
    \includegraphics[width=\linewidth]{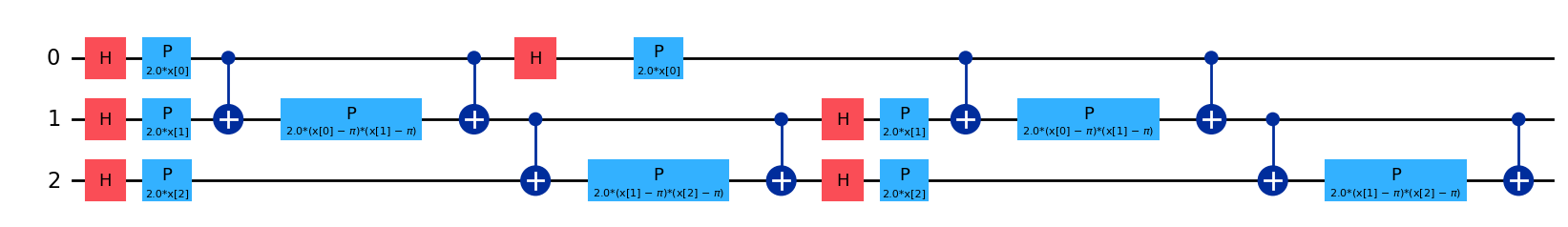}
    \subcaption{\texttt{ZZ} feature map repeated twice\label{fig:map_c}}
  \end{minipage}

  \caption{Examples of feature maps on two qubits.}
\end{figure}

To identify the optimal hyperparameters to find a quantum kernel, we evaluated several values for the following parameters:
\begin{description}
  \item[Number of qubits available:]
        Determines the upper limit of encodable features.
        When the number of qubits was insufficient, we applied Principal Component Analysis (PCA) to project the data into an \emph{n}-dimensional subspace capturing the most relevant variance.
        We evaluated the pipeline using 4, 8, 16, and 30 qubits, which represents the maximum number of features in the dataset.

  \item[Feature map:]
        The quantum circuit responsible for encoding classical data into quantum states.
        These circuits determine the geometry of the induced Hilbert space and affect model expressivity.
        Figure~\ref{fig:map_a}, Figure~\ref{fig:map_b} and Figure~\ref{fig:map_c} are three examples of possible feature maps.
        The feature maps considered in our experiments include \texttt{Z\_feature\_map}, \texttt{ZZ\_feature\_map}, \texttt{SU2HR}, and \texttt{SU2RR}.

  \item[Repetitions of the feature map:]
        Defines how many times the feature map circuit is applied in sequence.
        The reason repetitions of the same feature map may be required is that, in certain cases, a single application is not sufficient to capture the desired non-linear transformation $\phi(\vec{x})$.
        This layered construction is analogous to deep architectures in classical machine learning, where increasing depth enhances representational capacity.
        Figure~\ref{fig:map_c} shows a \texttt{ZZ\_feature\_map} repeated twice.
        We tested circuits with both one and two repetitions of the feature map.
\end{description}
To evaluate the quality of the kernel derived from the various feature maps, we employed the \textit{Kernel-Target Alignment} (KTA) strategy \cite{kta}.
This metric allows for estimating the effectiveness of a given kernel on a reference dataset without the need to train a machine learning model.

Formally, given a kernel matrix $K \in \mathbb{R}^{n \times n}$ computed over a set of $n$ examples, and a label vector $\mathbf{y} \in \{-1, 1\}^n$, the KTA is defined as:

$$\text{KTA}(K, \mathbf{y}) = \frac{\mathbf{y}^\top K\ \mathbf{y}}{\|K\|_F \cdot n}$$
where $\|K\|_F$ denotes the Frobenius norm of the kernel matrix.
This normalized inner product quantifies the alignment between the kernel matrix and the ideal target kernel implied by the labels.
A higher value of KTA indicates a stronger alignment and, consequently, a potentially more suitable kernel for classification tasks.

The top three configurations for the quantum kernel are reported in Table~\ref{tab:top_kernel}. All proposed configurations exhibit a high degree of alignment.
In this study, no single hyperparameter emerged as universally optimal across configurations.
This is evidenced by the presence of both circuits with many qubits and those with fewer qubits, deep circuits as well as shallow ones, and the use of three distinct feature maps.

\begin{table}
  \centering
  \caption{Top three kernel configurations.}
  \label{tab:top_kernel}
  \begin{tabular}{ccccr}
    \toprule
    \textbf{\# Qubits} & \textbf{Feature Map} & \textbf{Repetitions} & \textbf{Alignment (KTA)} \\
    \midrule
    30                 & SU2HR                & 1                    & 98.650\%                 \\
    8                  & ZMAP                 & 2                    & 93.558\%                 \\
    16                 & SU2RR                & 1                    & 91.203\%                 \\
    \bottomrule
  \end{tabular}
\end{table}

\subsection{Selecting the regularization parameter C for Quantum Annealing SVM}\label{sec:qasvm}

In our experiments, we explored a range of values for $C$ by selecting powers of two, from $2^2 - 1$ to $2^{12} - 1$.
This choice is motivated by the way optimization variables are encoded in our quantum annealing formulation of SVMs. Specifically, each continuous optimization variable is discretized and represented using $\log_2(C)$ binary variables.
By choosing $C$ as $2^b-1$, where $b$ is the number of bits used, we ensure that the integer values these binary variables can represent range from 0 to $C$, thereby fully utilizing all possible bit combinations.
This approach minimizes the number of unused configurations in the binary encoding and ensures a compact and efficient use of the available qubits, which is crucial given the hardware limitations of current quantum annealers.

We observed the best performance using regularization parameter values of $C = 7, 63, 255$.
Nevertheless, the differences in performance across the various configurations are minimal, suggesting that—in the application context we investigated—the examples are likely to be linearly separable with not excessive rate.

\subsection{Fully Quantum SVM}\label{sec:qkernel_and_qasvm}

By combining quantum kernels with optimization via quantum annealing, we designed a QSVM pipeline consisting of the following steps:
\begin{enumerate*}
  \item Computing the quantum kernel matrix $K$ using a gate-based quantum device, which encodes pairwise similarities between training examples in an implicit feature space;
  \item Formulating an optimization problem where $K$ is used to construct the objective function;
  \item Solving this problem using quantum annealing to determine the optimal model parameters;
  \item Using the obtained parameters to perform inference on new examples.
\end{enumerate*}
These models are designed to solve supervised classification tasks by learning a hyperplane that separates samples belonging to the positive class from those of the negative class.

The best-performing model was based on the kernel with the highest alignment (see Table~\ref{tab:top_kernel}), specifically the kernel constructed using 30 qubits and, among the candidate values introduced in Section~\ref{sec:qasvm}, a regularization parameter of $C=255$.
In this case, we achieved an \texttt{F1-score} of 90\%.
We also observed that the negative class exhibited higher \texttt{recall} but lower \texttt{precision}, whereas the opposite was true for the positive class.
This suggests that the trained model suffers from classification issues that lead to the generation of false negatives.
False negatives were found across all models, indicating that this is a systematic issue.
As usual, a more effective preprocessing phase would likely result in a uniform performance improvement across all configurations.

The model with the smallest number of qubits yielded the poorest results.
Using only 8 qubits to encode the input features and setting $C=63$, the resulting \texttt{F1-score} was 52\%, making this model comparable to random guessing.
This behavior may be attributed to an insufficient number of features used to represent the data: reducing from 30 to 8 qubits likely caused a significant loss of information, leading to examples becoming overly similar and thus harder to distinguish.

%---------------------
\section{Conclusion}

In this study, we have explored the feasibility and effectiveness of constructing an end-to-end quantum learning pipeline for support vector machines.
By integrating gate-based quantum kernel methods with quantum annealing-based optimization, we demonstrated that it is possible to train a fully quantum SVM model on real-world datasets.

Our results show that the choice of quantum kernel—particularly the number of qubits and the selected feature map—significantly influences classification performance.
Furthermore, we observed that the approximation involved in representing real-valued optimization variables with discrete binary variables in the QUBO formulation does not lead to substantial performance degradation.

The best-performing configuration, employing 30 qubits and a regularization parameter $C=255$, achieved an F1-score of 90\%, a result comparable to that obtained using a classical SVM with an RBF kernel, which achieved an F1-score of 91\%. The marginal performance difference may be attributed to the discretization of optimization variables in the quantum annealing process.
The presence of systematic false negatives suggests that further improvements could be achieved with a more refined data preprocessing pipeline.
While incorporating domain knowledge about the origin and nature of the training and test datasets could also be beneficial, we believe that the most effective improvements are likely to come from enhanced preprocessing strategies or the use of larger and more representative datasets

This work is a first step towards future investigations into scalable quantum machine learning pipelines and highlights the role of quantum high-performance computing in facilitating hybrid computational workflows across diverse quantum architectures. 

\bibliography{refs.bib}
\end{document}